\documentclass[11pt,letterpaper]{article}
\pdfoutput=1
\usepackage{jheppub}
\usepackage{pstricks}
\usepackage{skak}
\usepackage{wasysym}
\usepackage{graphicx, caption, subcaption}

\graphicspath{{./Figures/}}

\newcommand{\beq}{\begin{equation}}
\newcommand{\eeq}{\end{equation}}
\newcommand{\bea}{\begin{eqnarray}}
\newcommand{\eea}{\end{eqnarray}}
\newcommand{\be}{\begin{eqnarray}}
\newcommand{\ee}{\end{eqnarray}}

\def\tilde{\widetilde}
\def\hat{\widehat}

\def\CD{{\mathcal D}}

\def\CO{{\mathcal O}}

\newcommand{\pinf}{{\Omega}}

\newcommand{\dP}{{\tilde P}}
\newcommand{\dQ}{{\tilde Q}}
\newcommand{\dM}{{\tilde M}}
\newcommand{\deta}{{\tilde \eta}}
\newcommand{\dalpha}{{\tilde \alpha}}
\newcommand{\dbeta}{{\tilde \beta}}
\newcommand{\dmu}{{\tilde \mu}}
\newcommand{\dnu}{{\tilde \nu}}
\newcommand{\dplus}{{\tilde +}}
\newcommand{\dminus}{{\tilde -}}
\newcommand{\dpm}{{\tilde \pm}}

\setboardfontsize{8}

\title{Conformal constraints on defects}

\author{Abhijit Gadde}
\affiliation{Institute for Advanced Study, Princeton NJ 08540, USA}
\emailAdd{abhijit@ias.edu}

\abstract{
In this paper we study the constraints imposed by conformal invariance on extended objects \emph{a.k.a} defects in a conformal field theory. We identify a particularly nice class of defects that is closed under conformal transformations. Correlation function of the defect with a \emph{bulk} local operator is fixed by conformal invariance up to an overall constant. This gives rise to the notion of defect expansion, where the defect itself is expanded in terms of local operators. This expansion generalizes the idea of the boundary state. We will show how one can fix the correlation function of two defects from the knowledge of the defect expansion. The defect correlator admits a number of conformal cross-ratios depending on their dimensionality. We find the differential equation obeyed by the conformal block and solve them in certain special cases.
}


\begin{document}
\maketitle

\section{Defects in conformal field theories}
A conformal field theory is usually formulated in terms of local operators and their correlation functions. These correlation functions are strongly constrained by  conformal symmetry. In fact, the symmetry  fixes them completely modulo the discrete data associated to the three point functions known as operator product expansion coefficients. These data is the only dynamical information about the CFT and the rest is kinematics. Extended objects \emph{a.k.a} defects form an interesting class of operators in a CFT. They are important both from theoretical and experimental point of view. 
The most well-studied ones are the boundaries and lines in 2d CFTs \cite{Cardy:1984bb, Cardy:1989ir}, see \cite{Frohlich:2006ch} and references therein for defects in rational CFTs. Other examples include boundaries in higher dimensions \cite{McAvity:1995zd},  Wilson and 't Hooft line operators \cite{Shifman:1980ui, Berenstein:1998ij, Kapustin:2005py, Gomis:2009xg} which serve as order parameters for gauge theories, the monodromy defect of the $3d$ Ising model \cite{Billo:2013jda, Gaiotto:2013nva} and the twist operator which glues multiple copies of theories along a co-dimension $2$ locus \cite{Calabrese:2004eu, Hung:2014npa} etc.. Experimentally, any system at criticality is  in contact with its container. The associated boundary is a co-dimension $1$ defect. Moreover, introduction of an impurity in a critical system can engineer defect operators of various dimensions \emph{e.g.} colloids suspended in critical fluids such as oil-water mixture at critical temperature and concentration. See \cite{refId0} for an introduction to the experimental work and  references therein for the details. The ubiquitousness of defects in conformal field theories makes a convincing case for a systematic study of the constraints conformal symmetry imposes on their correlation functions. 

There has been a significant amount of work on BPS defect operators in supersymmetric conformal field theories. Such an operator is obtained either by introducing a singularity and/or a source in the path integral along its support or by introducing new degrees of freedom on the support, coupling them to the bulk and integrating them out. Although such a construction has been useful in many ways, it suffers from two serious drawbacks.
One, they need identification of \emph{weakly coupled}  fields of the microscopic theory and two, they completely mask defect correlations. We seek to overcome these drawbacks by studying defects from an abstract point of view. In the process, we will use a notion of that is similar to the operator product expansion wherein a defect itself is expanded into local operators \cite{Shifman:1980ui, Berenstein:1998ij, Gomis:2009xg}.
We show  that their two point function is fixed by conformal symmetry modulo \emph{defect expansion} coefficients, just as in the case of local operators.

\paragraph*{Note:}
While this paper was in preparation \cite{Billo:2016cpy} appeared which also deals with defects in conformal field theories and has a small overlap with the material  presented here.

\subsection{Conformal defects}
The conformal symmetry group is a group of transformations that keeps all the angles fixed. This also means that it takes the metric into itself up to an overall, possibly position dependent  factor. It is generated by the action of familiar Lorentz group along with special conformal transformations and scaling. In a $d$-dimensional Minkowski space, the conformal symmetry group is $SO(d,2)$ while in $d$ dimensional Euclidean space, it is $SO(d+1,1)$.
In this paper we will deal with conformal field theories in Euclidean space. The  space is compactified into a sphere $S^d$ by adding a point at infinity.
 
Using state operator correspondence any conformal field theory state on the sphere $S^{d-1}$ can be mapped to a local operator $\CO(x)$. The support of the local operator, i.e. the point $x$, is kept fixed by the subgroup $H_{\rm pt}:=SO(1,1)\times SO(d) \ltimes {\mathbb R}^d$. These factors correspond to scaling, rotations and special conformal transformations, about $x$, respectively. Standard arguments show that the local operator should furnish the representation of its stabilizer,  also known as the little group. One is usually interested in the finite dimensional representation of the little group. We take the action special conformal transformations to be vanishing and label the representation of the local operator by scaling dimension and spin. A representation of little group induces a representation of the full conformal group in a canonical fashion.
The space of all inequivalent local operator insertions, $SO(d,2)/H_{\rm pt}$, is isomorphic to $S^d$ as expected.

Extended operators also are an integral part of a conformal field theory.  In what follows, it is convenient to characterize them with their co-dimension rather than the dimension. A generic co-dimension $m$ defect does not preserve any symmetries. But if the the defect is translationally invariant and spans a flat hyperplane then it is clear that the support of such a defect is fixed by the subgroup $H_m:=SO(m)\times SO(d-m+1,1)$. This is the maximal subgroup that a co-dimension $m$ locus could preserve. 
The first factor is the rotations in the directions orthogonal to the defect and the second factor is the conformal transformation in the parallel directions. As we shall see in section \ref{embedding-defect}, the action of special conformal transformations in the orthogonal direction changes a flat defect into a spherical one. Clearly, the transformed defect also has the same stabilizer. 
The space of all co-dimension $m$ spheres is then locally given by the coset $SO(d+1,1)/H_m$.  The dimension of this space is $m(d-m+2)$.  The zero dimensional defect is supported on $S^0$, which is simply a pair of points. The space of such defects is $2\cdot d$ dimensional, in agreement with the above formula. Another interesting case is that of co-dimension $1$ defect. The dimension of its configuration space is given to be $d+1$. This is expected because a co-dimension one sphere is specified by its center ($d$-parameters) and radius.
The subgroup $SO(m)\subset H_m$ plays a special role as it preserves the defect support point-wise. This allows us to label the defect by its spin under $SO(m)$. We will mostly be working with scalar defects although our analysis can be generalized to the spinning defect as well.

Most studies of the defects in conformal field theory are from the point of view of their ``world-volume". In a CFT, in addition to the usual local operators there are local operators that are supported  only on the defect. We call such operators defect-local operators. They share many of the properties with the usual bulk-local operators. In particular they have a closed operator product expansion.  Because of the closure of the OPE, one can think of a given defect as supporting a conformal theory \emph{of its own}. Albeit the defect theory does have non-vanishing correlation with the bulk-local operators. 
Despite many parallels, the defect theory is different from the usual conformal field theory in one crucial aspect: it does not have a stress tensor. This is to be expected because the defect system freely exchanges energy with the bulk.   As a result, the Ward identity for stress tensor gets modified in the presence of the defect. For the case of a flat defect, the Ward identity is
\be\label{displacement}
\partial_i T^{i \alpha}(x)= D^\alpha(x)\delta_D,\qquad i=1,\ldots, d,\qquad \alpha=1,\ldots, m.
\ee
Here $\alpha$ labels the directions orthogonal to the defect. The $\delta_D$ is a delta function supported at the defect and $D^\alpha(x)$ is the defect-local operator which displaces the defect at $x$ in the transverse direction $\alpha$. 
Our viewpoint is going to be slightly different from the above. We are interested in considering the correlation functions of multiple defects. In a broad sense, our approach could be thought of as a second quantized formalism for defects. 

The outline of the paper is as follows. In section \ref{embedding-space}, we will introduce the embedding space formalism which realizes conformal symmetry  linearly. We will use the embedding space formalism to construct the so called conformal defects. Such a construction has a benefit of making all the symmetries manifest. In section \ref{defect-expansion}, we will show that the form of the two point function of a spherical defect with a bulk-local operator is uniquely fixed by conformal symmetry. This is used to define a sort of operator product expansion in which the spherical defect is expanded in terms of bulk-local operators. The coefficients of the expansion are the dynamic data of the defect. Correlation function of two defects are studied in section \ref{two-defect}. Configurations of two defects of generic dimensions admit a number of generic conformal cross-ratios. We obtain partial differential equations obeyed by the associated conformal blocks and solve them in some cases. In section \ref{examples}, we present a discussion about the scope and generalization of our approach.

\section{Linear realization of conformal symmetry}\label{embedding-space}
A Euclidean conformal field theory in $d$ dimensions is invariant under
\begin{eqnarray}
{\rm Translation:} && x^i \to x^i + a^i \\
{\rm Rotation:}  &&  x^i \to M^i_{\, j} \, x^j\\
{\rm Dilation:}&&  x^i \to \lambda \, x^i \label{scaling}\\
{\rm Special \,conformal \,transformation:}  && x^i \to\frac{x^i  +b^i x^2}{1 +2 b\cdot x +b^2 x^2}
\end{eqnarray}
where $M^i_{\, j}$ is a $SO(d)$ matrix. These transformations together generate the symmetry group $SO(d+1,1)$. This is not quite manifest, especially because the translations and the special conformal transformations act in a nonlinear way on the coordinates. The form of the symmetry group suggests that there is a linear realization of the $d$-dimensional conformal symmetry on the bigger space ${\mathbb R}^{d+1,1}$. This is indeed so \cite{Dirac:1936fq, Boulware:1970ty, Ferrara:1973eg, Weinberg:2010fx}. In this context, ${\mathbb R}^{d+1,1}$ is known as the embedding space. More recently, the embedding space formalism has been applied to compute four point conformal blocks for spinning operators \cite{Costa:2011mg, SimmonsDuffin:2012uy}.
We reserve upper case letters $X, Y$ etc. to denote  its coordinates. 
In order to go from the embedding space to the original $d$-dimensional space, we need to get rid of two dimensions all the while preserving the action of $SO(d+1,1)$. This is achieved by restricting to the projective null cone i.e. to the points $X$ of the embedding space satisfying $X^2=0$ up to the  identification $X\sim g X$ (for $g \in {\mathbb R}$). This $GL(1)$  gauge redundancy should not be confused with the \emph{physical} scale transformation \eqref{scaling}.

Let us see this in more detail.  It is convenient to use the light-cone coordinates $X^A=(X^+,X^-, X^i)$,  $i=1,\ldots, d$ for the embedding space. The $SO(d+1,1)$ invariant dot product is defined as
\be
X\cdot Y=-\frac{X^+Y^-+X^-Y^+}{2}+X^i Y^i.
\ee
Null condition $X^2=0$ is solved by the vector $X=(\alpha,x^2/\alpha, x^i)$ where $x^2=x^ix^i$. The projectivization $X\sim g X$ is taken care of by fixing the gauge   $X^+=1$.  This gauge  condition is called the Poincare section. It allows us to identify  points on  the Poincare section  $(1,x^2, x^i)$ with the points $x^i$ in the original space. Alternatively stated, a point $x^i$ has a unique lift to the Poincare section of the embedding space, namely $(1,x^2, x^i)$. The only point absent from the Poincare section is the point at infinity. We normalize its lift to the embedding space as $\pinf=(0,1,0^\mu)$.
The linearity of embedding space is reflected in a rather useful fact that the distance squared  between two points $x^i$ and $y^i$ is given by $-2X\cdot Y$ where $X$ and $Y$ are the Poincare section lifts of $x$ and $y$ respectively. 

A scalar local operator $\phi(x)$ of conformal dimension $\Delta$ has a lift $\Phi(X)$ to the embedding space null cone. In order for the $SO(d+1,1)$ transformations to respect the Poincare section, $\Phi(X)$ should actually be function of the ratio $X^\mu/X^+$. Taking into account the scaling property, we define $\Phi(X)\equiv (X^+)^{-\Delta}\phi(X^\mu/X^+)$. The effectiveness of the embedding space formalism is apparent when we consider two point function $\langle \Phi(X)\Phi(Y)\rangle$. It should be a homogeneous scalar function of $X$ and $Y$ with degree $\Delta$ in both. There is a unique choice,
\be\label{local2point}
\langle \Phi(X)\Phi(Y)=\frac{1}{(-2 X\cdot Y)^\Delta}.
\ee
In our conventions, local operators are normalized such that their two point function always has the above form.


\subsection{Defects in embedding space}\label{embedding-defect}

\begin{figure}[h]
\centering
\includegraphics[scale=0.3]{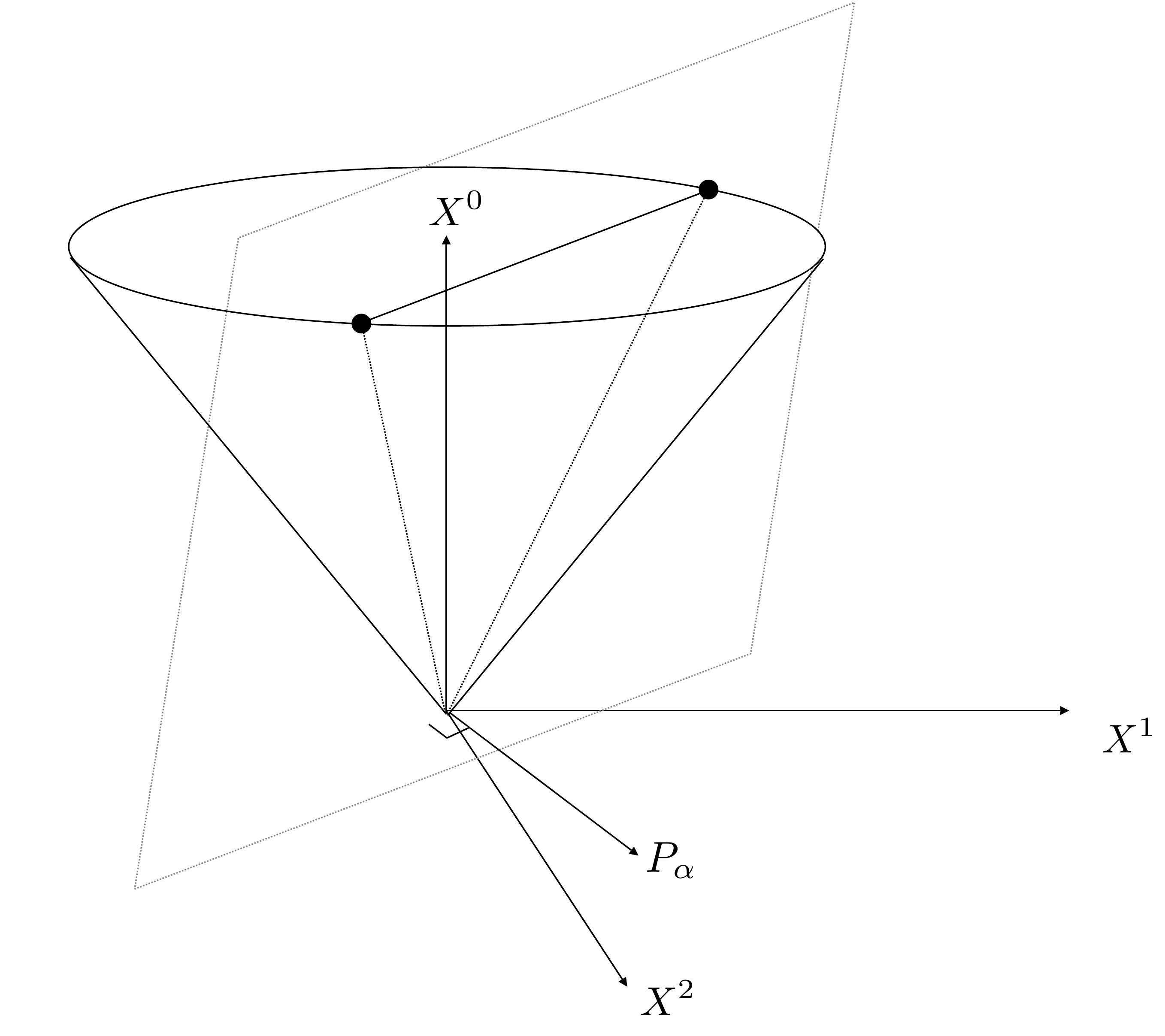}
\caption{Null cone in the $3$-dimensional embedding space and its intersection with a $2$-hyperplane resulting in a spherical $0$-defect (i.e. pair of points, denoted by solid dots). The orthogonal vectors $P_\alpha$ parametrize the hyperplane and hence the defect.}
\label{nullcone}
\end{figure}

As discussed earlier, we are interested in co-dimension $m$ defects whose support preserves $SO(m)\times SO(d-m+1,1)$. The best way to identify such a locus is to analyze this condition in the embedding space. 
A hyperplane in ${\mathbb R}^{d+1,1}$ is classified as time-like, if it intersects the null cone;  space-like, if it does not intersect the null cone and light-like, if it is tangent to the null cone.
It is clear that a co-dimension $m$ time-like hyperplane precisely preserves the subgroup in question. The intersection of such a hyperplane with the null-cone is $d-m+1$ dimensional and projectivization reduces one dimension further giving an $d-m$-dimensional locus in the orginal space. As we will show shortly, this locus is a sphere. The idea is illustrated in figure \ref{nullcone} for the case of a $0$-dimensional sphere.

Let us characterize the co-dimension $m$ hyperplane by specifying $m$ transverse vectors $P_\alpha$ ($\alpha=1,\ldots, m$). The defect locus is $X$ satisfying
\be\label{defect-locus}
X\cdot X=0,\quad P_\alpha\cdot X=0
\ee
up to the projective identification. Note that unlike $X$, vectors $P_\alpha$ are not  null. 
Let us denote a defect supported  as such by $\CD^{(m)}(P_\alpha)$. Of course, the definition of a defect requires much more than specifying its support; by $P_\alpha$ we are simply referencing to its ``co-ordinate" and the superscript $(m)$ denotes its co-dimension.
Let us start by computing the image of a co-dimension one defect $\CD^{(1)}(P)$ in $d$-dimensional  space. Imposing the condition \eqref{defect-locus} on the null vector in the Poincare section $X =(1 ,x^2,x^i)$, we get
\be\label{codim1}
x^2+\frac{P^-}{P^+} -\frac{2 p\cdot x}{P^+}=0
\ee
This is the equation of a co-dimension one sphere centered at $p^\mu/P^+$ of radius $|P|/P^+$. 

In the case of $\CD^{(m)}(P_\alpha)$, it is clear that each of the $m$ vectors $P_\alpha$ gives rise to a co-dimension one sphere. When the plane transverse to $P_\alpha$ is time-like, these spheres have a non-empty intersection which is precisely a co-dimension $m$ sphere. This approach of thinking about the  sphere using $P_\alpha$ has an extra benefit. A new set of vectors $P'_\alpha$ obtained from $P_\alpha$ by a $GL(m)$ transformation also labels the same hyperplane and hence the same co-dimension $m$ sphere. In ${\mathbb R}^d$ this corresponds to many ways in which one can obtain the same co-dimension $m$-sphere as the intersection of different sets of $m$ co-dimension one spheres. Interestingly, invariance under this new extended $GL(m)$ gauge symmetry,  uniquely determines its center and radius. 

Again the case of co-dimension one sphere serves as a guiding example. We want to  express its the center and radius  in a $GL(1)$ invariant fashion. The notion of distance is not absolute in conformal field theory. It is defined relative to something. It is natural to take the point at infinity $\pinf$ as the reference. Then the only nontrivial $GL(1)$ invariant null vector made out of $P$ and $\pinf$ is
\be\label{1-center}
C=\frac{(P\cdot P) \pinf-2(P\cdot \pinf) P}{4(P\cdot \pinf)^2}.
\ee
We have normalized it so that it belongs to the Poincare section. This must be the center. Compared to \eqref{codim1} we see that it is indeed the case. The radius is the distance between the center and a generic point on the defect. Its square is computed by $-2C\cdot X$ for any point $X$ on the sphere 
\be\label{1-radius}
r^2=-2 C\cdot X=\frac{(P\cdot P)}{4(P\cdot \pinf)^2}.
\ee
We have used $P\cdot X=0$. This also agrees with the explicit equation \eqref{codim1}.

For a spherical defect $\CD^{(m)}(P_\alpha)$, the problem of finding the center and radius becomes that of constructing $GL(m)$ invariant expressions of $P_\alpha$ and $\pinf$.
Instead of doing it directly, it helps to gauge fix $P_\alpha\cdot P_\beta=\delta_{\alpha \beta}$. In this gauge the defect is characterized not by $m$ arbitrary transverse vectors but rather by  an $m$-dimensional orthonormal frame. The gauge fixed ``coordinates" still have a remnant $O(m)$ gauge redundancy. Now it is easy to construct $O(m)$ invariants out of orthonormal $P_\alpha$'s and $\pinf$.
\be\label{center-radius-ortho}
C=\frac{\pinf-2(P_\alpha\cdot \pinf) P_\alpha}{4(P_\gamma\cdot \pinf)(P_\gamma\cdot \pinf)},\qquad
r^2=\frac{1}{4(P_\gamma\cdot \pinf)(P_\gamma\cdot \pinf)}.
\ee
We will always work in this gauge.

Given a sphere in the $d$-dimensional space, how to find its coordinates $P_\alpha$?
By definition, the vectors $P_\alpha$ are orthogonal to the $d-m+2$ dimensional hyperplane in the embedding space. Fixing such a hyperplane needs specification of $d-m+2$ number of vectors. Coincidentally, to fix a co-dimension $m$-sphere i.e. a $d-m$ dimensional sphere in ${\mathbb R}^d$ we also need to pick the same number of points. Using this observation,  coordinates $P_\alpha$ of a given $\CD^{(m)}$ can be determined as follows.
Pick any $d-m+2$ points on $\CD^{(m)}$ and consider their lifts to the Poincare section $X_k, \, k=1,\ldots, d-m+2$. The coordinate vectors $P_\alpha$ simply span the space of solutions to $X_k\cdot P=0$.
Let us illustrate this for a few examples. 
\begin{itemize}
\item
Consider a defect of radius $r$ centered at the origin. The defect is aligned so that it lies in the $d-m+1$ dimensional plane spanned by orthonormal  basis vectors $e_j, j=1,\ldots d-m+1$. We pick $X_j=(1,r^2, r e_j)$ and $X_{d-m+2}=(1,r^2, -re_1)$. 
A convenient orthonormal basis satisfying $X_k\cdot P=0$ is 
\be\label{sphere-p}
P_\alpha=(0,0,e_{d-m+1+\alpha}), \,{\rm for}\,\alpha=1,\ldots, m-1,\qquad\qquad P_{m}=(\frac1r, -r,0). 
\ee
Substituting them in equation \eqref{center-radius-ortho} produces the expected result. 
\item
If the defect shifted along $e_1$ by distance $\ell$, we pick the points $X_j=(1,r^2, r e_j+\ell e_1)$ and $X_{d-m+2}=(1,r^2, -re_1+\ell e_1)$ on it. The coordinates $P_\alpha$'s satisfying $X_k\cdot P=0$ is
\be\label{shifted-defect}
P_\alpha=(0,0,e_{d-m+1+\alpha}), \,{\rm for}\,\alpha=1,\ldots, m-1,\qquad\quad P_{m}=(\frac1r,-r+\frac{\ell^2}{r},\frac{\ell}{r} e_1). \qquad
\ee
\item
Now let consider a flat defect aligned in a plane spanned by $e_j, j=1,\ldots d-m$. We pick the points $X_j=(1,1,e_j), (1,0,\vec 0), \pinf$ on it. Then we have
\be\label{flat-defect}
P_\alpha=(0,0,e_{d-m+\alpha}), \,{\rm for}\,\alpha=1,\ldots, m.
\ee
\item
If the flat defect is tilted by angle $\theta$ in $e_1-e_d$ plane, then the coordinates become
\be\label{tilted-defect}
P_\alpha=(0,0,e_{d-m+\alpha}), \,{\rm for}\,\alpha=1,\ldots, m-1, \qquad \quad P_m=(0,0,\cos \theta e_d-\sin\theta e_1).\qquad
\ee
\end{itemize}

Alternatively, the orthonormal frame transverse to $P_\alpha$'s can also be used to parametrize the defect. We take this frame to be spanned by vectors $\dP_{\dalpha}, \,\dalpha=1,\ldots, d-m+2$. As this ``dual" frame is time-like, the orthonormality condition is $\dP_{\dalpha}\cdot \dP_{\dbeta}=\eta_{\dalpha\dbeta}$ where $\eta_{\dalpha\dbeta}$ is the flat Minkowski metric. Unless otherwise mentioned, we will stick to the parametrization with $P_\alpha$'s.

\section{Defect expansion}\label{defect-expansion}

In this section we discuss the correlation function of the conformal defect with a bulk-local operator.  The symmetry preserved by a defect $\CD^{(m)}$ is $SO(d+1-m,1)\times SO(m)$. With the addition of a bulk local operator $\Phi$ this symmetry is broken to $SO(d+1-m)\times SO(m-1)$. A good  way to see this is to conformally map the defect configuration to $AdS_{d+1-m}\times S^{m-1}$ and insert the local operator at the origin of $AdS$. 
As discussed earlier, $\CD^{(m)}$ can have spin under the $SO(m)$ transverse rotation and  is neutral under $SO(d+1-m)\subset SO(d+1-m,1)$. The local operator $\Phi$, on the other hand, transforms under the whole $SO(d)$ rotational symmetry. The correlation function is non-zero only if the configuration is invariant under the preserved symmetry group $SO(d+1-m)\times SO(m-1)$ \emph{i.e.} if we let the defect and local operator transform as $R_\CD$ and $R_\Phi$ under this group, we expect a nonzero correlation only if $R_{\CD}\otimes R_{\Phi}$ contains a singlet. In what follows, we stick to defects that transform trivially under $SO(m)$. 

Let us see what this means for $\CD^{(1)}$. To have a non-vanishing correlation, the local operator should be a singlet under the whole $SO(d)$ \emph{i.e.} it should be a scalar operator. For the case of $\CD^{(2)}$, we expect the local operator to be a singlet under $SO(1)\times SO(d-1)$. As a result, it can only be in a traceless symmetric tensor. The same analysis holds for $\CD^{(d)}$ defect \emph{i.e.} a pair of scalar local operators.  Higher co-dimension defects can have correlation with local operators transforming in more complicated representations but to keep the analysis simple, we only study correlation with scalar local operators.


\subsection{Correlation with local operators}
We normalize the defect so that its one point function is $1$.
The form of the two point function of the defect and a local operator is completely fixed by conformal invariance. Let $\Delta_\Phi$ be the conformal dimension of the  scalar local operator $\Phi(X)$. As the vectors $P_\alpha$ have an $O(m)$ gauge redundancy, we require the correlation function to be invariant under $O(m)$ as well.
The only conformally invariant and gauge invariant two point function that has appropriate scaling with respect to $X$ is
\be\label{defect-bulk}
\langle \CD^{(m)}(P_\alpha) \Phi(X)\rangle=C^{\CD}_{\Phi} 
\Big((P_\gamma\cdot X)(P_\gamma\cdot X)\Big)^{-\frac{\Delta_\Phi}{2}}.
\ee
As the defect and the local operator have been separately normalized, the  coefficient $C^{\CD}_{\Phi}$ is a physically meaningful parameter. Let us analyze the form of this two point function in more detail.
\begin{figure}[h]
\centering
\includegraphics[scale=0.35]{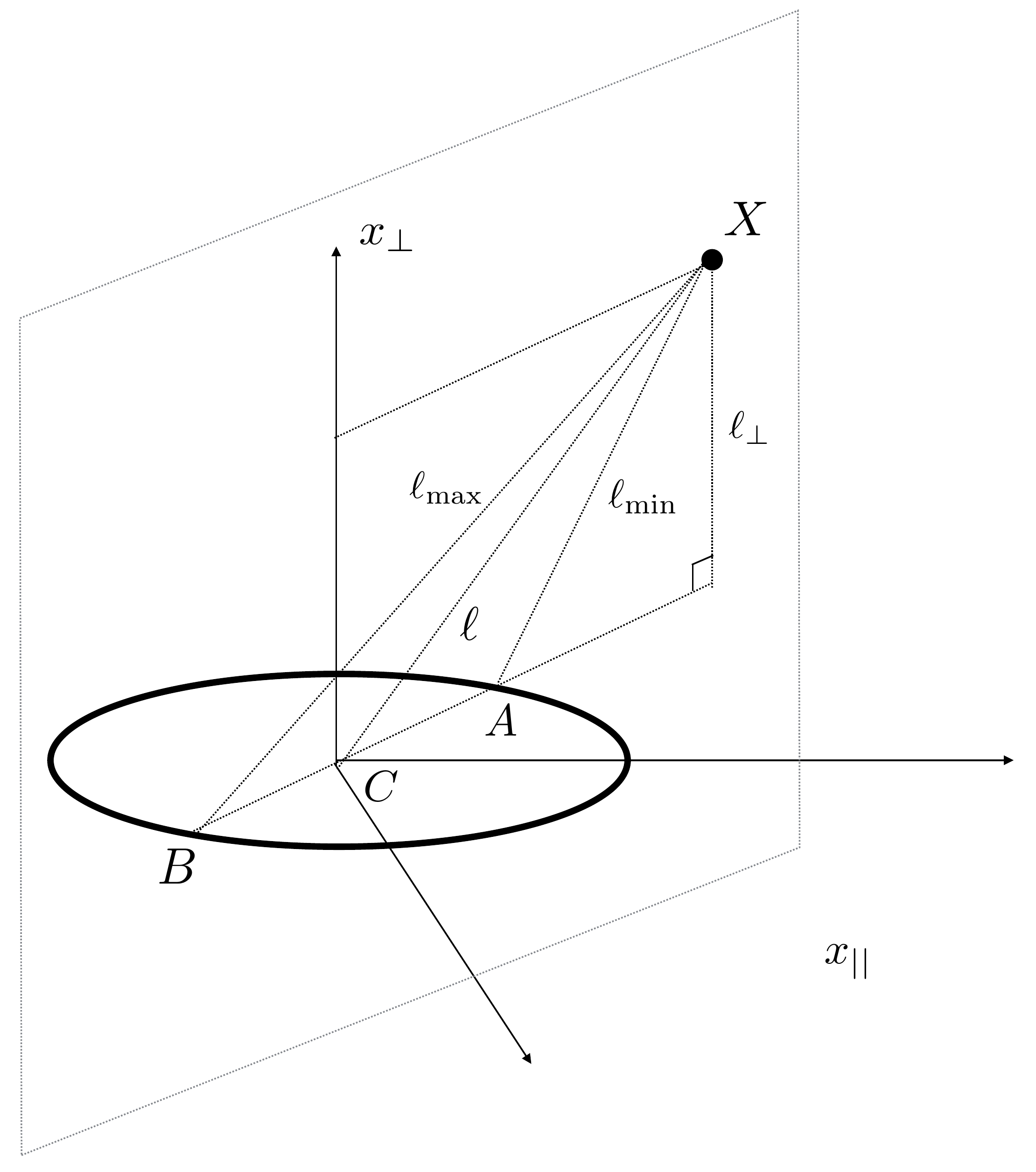}
\caption{A generic configuration of a circular defect and bulk-local operator.}
\label{defect-localop}
\end{figure}

Let the defect $\CD^{(m)}$ be of radius and centered at the origin.  Align it so that it lies in the $d-m+1$ dimensional plane spanned by orthonormal  basis vectors $e_i, i=1,\ldots d-m+1$. Let the local operator $\Phi$ be at a generic position $X$. This configuration is shown in figure \ref{defect-localop}. The coordinates $P_\alpha$ of such a sphere have been determined in equation \eqref{sphere-p}.
Substituting them in equation \eqref{defect-bulk}, we get
\be\label{explicit-defect-bulk}
\langle \CD^{(m)}(P_\alpha) \Phi(X)\rangle=C^{\CD}_{\Phi} \Big(\ell_{\perp}^2+\frac14(\frac{\ell^2}{r}-r)^2\Big)^{-\frac{\Delta_\Phi}{2}}
=C^{\CD}_{\Phi}  \Big(\frac{\ell_{\rm min} \ell_{\rm max}}{2r}\Big)^{-\Delta_\Phi}.
\ee
Here $\ell_{\perp}^2=\sum_{\alpha=1}^{m-1} x_{d-m+1+\alpha}^2$ is the perpendicular distance squared of the local operator from the $d-m+1$-dimensional hyperplane \emph{containing} the defect. Length $\ell_{\rm min}$ ($\ell_{\rm max}$) is the minimum (maximum) distance of the local operator from the defect.

\subsection*{Planar limit}
A special case of interest is where the defect is flat and spans an $d-m$-dimensional hyperplane. It is obtained from a generic spherical defect by taking the radius to infinity. This limit is easily taken in equation  \eqref{explicit-defect-bulk}.  As $r\to \infty$ limit with $\ell_{\rm min}$ fixed,  $\ell_{\rm max}\to 2r$. Equation \eqref{explicit-defect-bulk} reduces to
\be\label{planar-bulk}
\langle\CD^{(m)}(P_\alpha)\Phi(X)\rangle=C^{\CD}_{\Phi} \ell_{\rm min}^{-\Delta_\Phi}.
\ee
We can get the same result by using the coordinates $P_\alpha$ the flat defect explicitly \eqref{flat-defect}.

\subsection*{Additional insertion of defect-local operator}
The correlation function of the defect with a a defect-local operator  and bulk local operator is also determined by conformal invariance up to an overall constant. In addition to the  defect  $\CD^{(m)}(P_\alpha)$ and bulk operator $\Phi(X)$, let us consider the insertion of a scalar operator $o(Y)$ of conformal dimension $\Delta_o$ on the defect. The coordinate $Y$ labels a point on the defect, hence it obeys $P_\alpha\cdot Y=0$. Their  correlation function is uniquely fixed
\be
\langle \CD^{(m)}(P_\alpha) o(Y) \Phi(X)\rangle=C^{\CD}_{\Phi, o} 
\Big((P_\gamma\cdot X)(P_\gamma\cdot X)\Big)^{\frac{\Delta_o-\Delta_\Phi}{2}}
(-2 X\cdot Y)^{-\Delta_o}.
\ee
For a planar defect oriented as before and $o(Y)$ inserted at the origin, the three point function simplifies.
\be
\langle \CD^{(m)}(P_\alpha) o(Y) \Phi(X)\rangle= C^{\CD}_{\Phi, o} |\ell|^{-2\Delta_o}|\ell_{\rm min}|^{\Delta_o-\Delta_\Phi}.
\ee
Here $|\ell|$ is the distance of the bulk operator from the origin and $\ell_{\rm min}$ is its minimum distance from the defect. Note that $C^\CD_{\Phi}=C^\CD_{\Phi,{\bf I}}$.

\subsection{Operator product expansion}\label{defectexp}
\begin{figure}[h]
\centering
\includegraphics[scale=0.3]{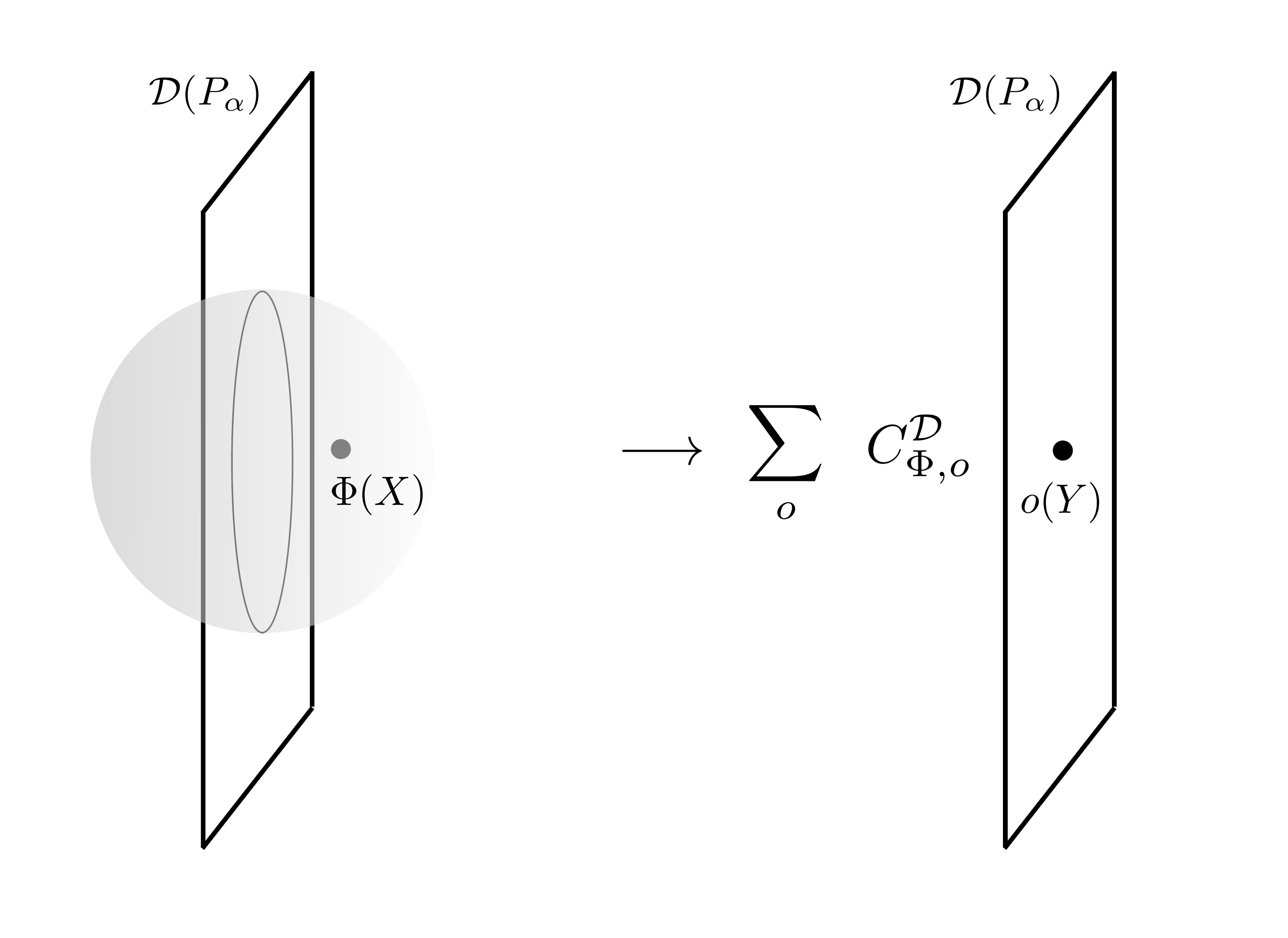}
\caption{A spherical slice enclosing the bulk operator $\Phi(X)$ and cutting a $2$-dimensional defect $\CD(P_\alpha)$. The state induces on the slice can be expanded in terms of defect local operators $o(Y)$.}
\label{ope}
\end{figure}

The correlation functions discussed so far can be used to define two notions of operator product expansion. 
Consider a bulk operator $\Phi(X)$ inserted near a defect $\CD(P_\alpha)$. Pick a spherical  slice for quantization that encloses the insertion point of the bulk operator and cuts through the defect. The bulk local operator induces a particular state on the spherical slice. Note that this state does not belong to the Hilbert space of the theory on the sphere but rather to the Hilbert space of the theory on a ``decorated" sphere where the decoration is provided by the intersection of the defect with the quantization slice. Using scale transformation, the decorated spherical slice can be scaled  to a point  \emph{on the defect}. In this way, a bulk operator inserted close to a defect can be expanded in terms of defect-local operators.
This is illustrated in figure \ref{ope}. In this expansion, the coefficient ${C}^\CD_{\Phi,o}$ can be thought of as the strength with which the bulk operator $\Phi$ induces the defect-local operator $o$ on the defect $\CD$.

The other notion of operator product expansion  arises when we consider the configuration in figure \ref{defectexp}.
\begin{figure}[h]
\centering
\includegraphics[scale=0.3]{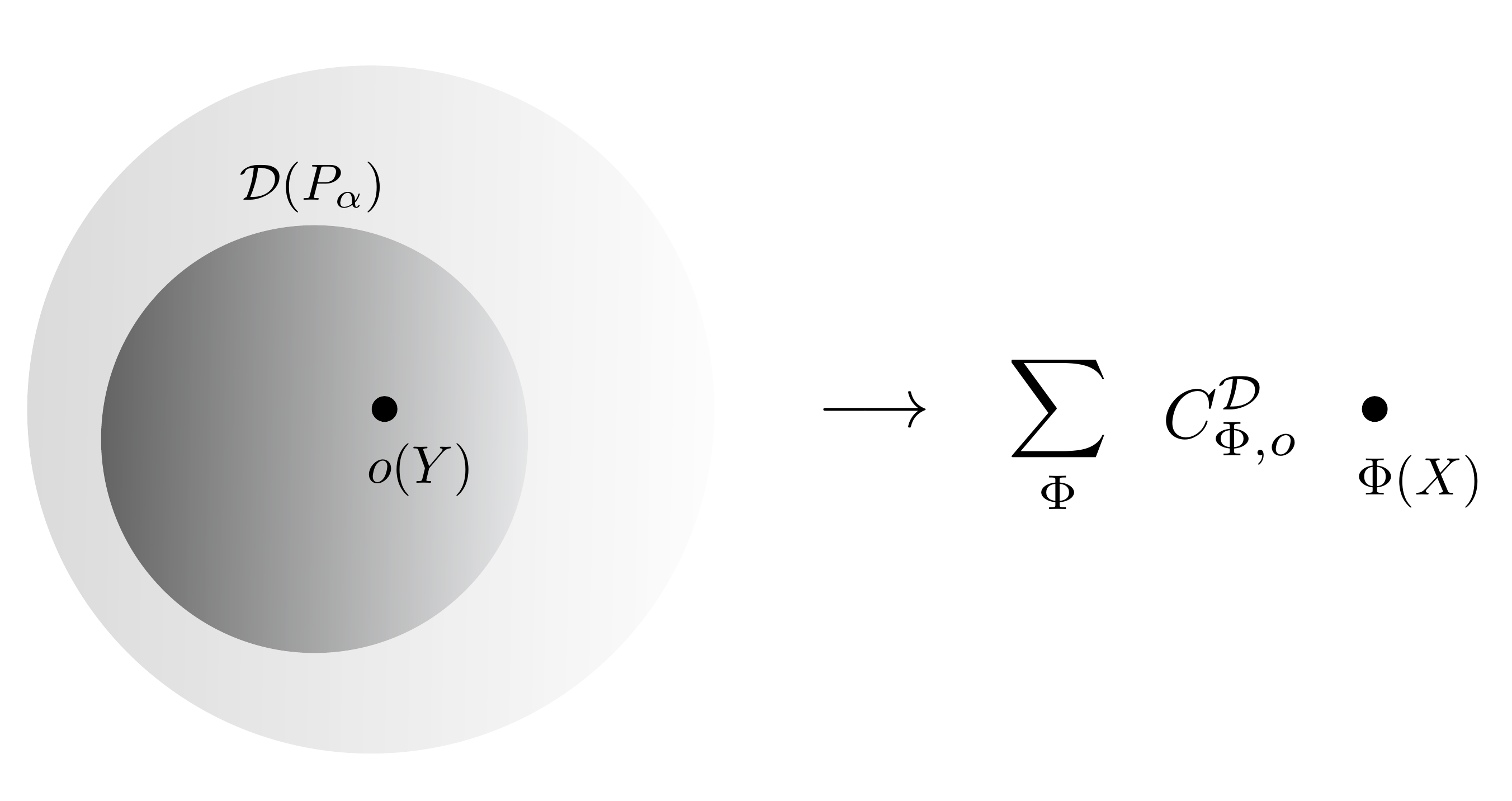}
\caption{The quantization slice encloses the defect. The state induced on the sphere is expanded in terms of bulk local operators.}
\label{defectexp}
\end{figure}
Instead of having the quantization slice cut the defect, we take it to enclose the defect   $\CD(P_\alpha)$ (and the defect-local insertion $o(Y)$). The state induced on the sphere can now be expanded in terms of bulk-local operators $\Phi(X)$ in the usual way. In this expansion, the coefficient $C^\CD_{\Phi,o}$ is the strength with which the defect $\CD$ with the insertion $o$ induce the bulk operator $\Phi$. 

It is the later expansion that we would mostly interested in. To distinguish it from the former, we call it the defect expansion. A defect without any defect-local operator $o$ can also be expanded in bulk operators $\Phi$. In this case, the relevant expansion coefficients are $C^\CD_\Phi=C^\CD_{\Phi,{\bf I}}$.
The defect expansion takes the form
\be
\CD(P_\alpha) =\sum_\Phi C^\CD_{\Phi}  \,\,f_{\Delta_\Phi}(P_\alpha, X, \partial_X) \Phi(X)
\ee 
Here $f_{\Delta_\Phi}(P_\alpha, X, \partial_X)$ is a differential operator fixed by demanding  that the correlation function of the defect with a probe operator $\Phi'(X')$ has the expected form \eqref{defect-bulk} \emph{i.e.}
\be
f_{\Delta_\Phi}(P_\alpha, X, \partial_X)&\,\,&\Big(4(X\cdot X')^2\Big)^{-\frac{\Delta_\Phi}{2}}=\Big((P_\alpha\cdot X')(P_\alpha\cdot X')\Big)^{-\frac{\Delta_\Phi}{2}}
\ee
The state induced by the defect is not an arbitrary one. Similar to the boundary state, it encodes the symmetries preserved by the defect. The contribution of the conformal multiplet of $\Phi(X)$ to the defect expansion is a generalization of the Ishibashi state.

\section{Two point function of defects}\label{two-defect}

In this section, we study the correlation function of two defects of arbitrary co-dimension, $\CD^{(m)}(P_\alpha)$ and $\CD^{(k)}(Q_{\rho})$. The indices $\alpha=1,\ldots, m$ and $\rho=1,\ldots, k$. Depending on $m,k$ and the dimension of space  $d$, configurations of the two defects admit a number of conformal cross-ratios. They  are combinations of $P_\alpha$ and $Q_\rho$ that are conformally invariant. The orthonormal frame coordinates $P_\alpha$ and $Q_\rho$ have $O(m)$ and $O(k)$ gauge redundancy. So, in addition to being conformally invariant, we need the cross-ratios to be gauge invariant as well. A simple example of a cross-ratio is
\be
\eta=(P_\alpha\cdot Q_\rho)(Q_\rho \cdot P_\alpha).
\ee
Let us now enumerate the number of cross-ratios admitted by $D^{(m)}$ and $D^{(k)}$. This turns out to be the easiest in the embedding space. 
There, the cross-ratios encode the  configurations of co-dimension $m$ and $k$ hyperplanes up to overall rotations. For example, if we are interested in cross-ratios of two circular defects in $2$-dimensions, we would be counting the parameters of two co-dimension one hyperplanes in $4$-dimensions. A such a hyperplane is uniquely specified by giving its normal  vector. So we might as well count the parameters labelling the relative configuration of two vectors. It is easy to see that the only such parameter is the angle between them. This tells us that the two circular defects admit only one conformal cross-ratio. In fact this is the case for two co-dimension $1$ defects in any dimension. In this case, the cross-ratio can be understood as follows.  Using conformal transformations, the co-dimension $1$ defects can be arranged to be concentric. Then the unique cross-ratio is essentially  the ratio of their radii. We will make this precise in section \ref{codim1}.

For general defects, conformal cross-ratios are enumerated as follows. Consider the case when the hyperplanes spanned by $P_\alpha$ and $Q_\rho$ generically only intersect at the origin. This happens when $m+k\leq (d+2)$.  Without loss of generality, let us  assume $m\geq k$.
The angles between the two hyperplanes are encoded in the $m\times k$ matrix $P_\alpha \cdot Q_\rho$ modulo residual gauge transformations. The gauge transformations can be used to simplify the matrix as follows. 
\bea
{\scriptstyle k}
\left\{
\begin{array}{c}
\\
\\
\\
\end{array}
\right.
\overbrace{
 \left[
\begin{array}{ccccc}
*\quad & *\quad & *\quad & *\quad & *\\
*\quad & *\quad & *\quad & *\quad & *\\
*\quad & *\quad & *\quad & *\quad & *
\end{array}
\right] 
}^{m}
\xrightarrow{SO(k)}&&
 \left[
\begin{array}{ccccc}
*\quad & *\quad & *\quad & *\quad & *\\
0\quad & *\quad & *\quad & *\quad & *\\
0\quad & 0\quad & *\quad & *\quad & *
\end{array}
\right]
\nonumber\\
&&\qquad\qquad\,\,\,
\Big\downarrow {\scriptstyle SO(m)}
\nonumber\\
&&
\left[
\begin{array}{ccccc}
*\quad & 0\quad & 0\quad & 0\quad & 0\\
0\quad & *\quad & 0\quad & 0\quad & 0\\
0\quad & 0\quad & *\quad & 0\quad & 0
\end{array}
\right]
\nonumber\\\nonumber
\eea
We have illustrated the case of $k=3,m=5$ with $d\geq 6$. The only gauge invariant data are the entries along the diagonal.
Moreover, we see that the $SO(m)$ action could have set yet another entry to zero. This corresponds to the unbroken conformal Killing vector. So we learn that for $d\geq m+k-2$, the number of conformal cross-ratios is ${\rm min}(m,k)$. The number of conformal Killing vectors is $|m-k|-1$ for $m\neq k$ and $0$ for $m=k$.

In the other case, when $d<m+k-2$, we can carry out the same analysis except for an important difference. The hyperplanes spanned by $P$ and $Q$ generically intersect in a $(m+k)-(d+2)$ dimensional hyperplane.  As a result the rectangular matrix $P_\alpha\cdot Q_\rho$ has a diagonal block of size $(m+k)-(d+2)$ with unit entries along its diagonal. Consider the previous example $k=3,m=5$ but with $d=5$.
\be
{\scriptstyle k}
\left\{
\begin{array}{c}
\\
\\
\\
\end{array}
\right.
\begin{array}{c}
{\scriptstyle (m+k)-(d+2)}
\left\{ \right.\\
\\
\\
\end{array}
\overbrace{
 \left[
\begin{array}{ccccc}
1\quad & 0\quad & 0\quad & 0\quad & 0\\
0\quad & *\quad & *\quad & *\quad & *\\
0\quad & *\quad & *\quad & *\quad & *
\end{array}
\right] 
}^{m}
\xrightarrow[SO(m)]{SO(k)}
 \left[
\begin{array}{ccccc}
1\quad & 0\quad & 0\quad & 0\quad & 0\\
0\quad & *\quad & 0\quad & 0\quad & 0\\
0\quad & 0\quad & *\quad & 0\quad & 0
\end{array}
\right]
\nonumber\\
\nonumber
\ee
The number of cross-ratios in this case are $k-(m+k-d-2)=d-m+2$. More generally it is ${\rm min}(d-m+2,d-k+2)$. 
We can understand this result in another way. In the case $d<m+k-2$, the $d-m+2$ and $d-k+2$ dimensional hyperplanes spanned by ``dual" frames $\dP$ and $\dQ$ generically only intersect at the origin. Applying earlier analysis to them, we immediately get the desired  result. 

Combining the two cases, we conclude
\bea\label{cross-ratio}
{\rm \#\, cross\mbox{-}ratios}&=&{\rm min}(m,k,d-m+2,d-k+2)\\
{\rm \#\, conformal \, Killing\, vectors}&=&
\left\{\begin{array}{cc}
|m-k|-1 &\qquad{\rm for}\quad m\neq k\\
0& \qquad{\rm for}\quad m=k.
\end{array}
\right.
\eea
As a sanity check, consider two $0$-defects \emph{i.e.} 4 points, in $d$-dimensions. From equation \eqref{cross-ratio} we see that the number of conformal cross-ratios is $2$ which agrees with our expectation. Also, for two co-dimension $1$ defects, we get a single cross-ratio as expected.

Because the remnant gauge symmetry in the orthonormal gauge is $O(m)\times O(k)$, the conformal cross-ratios are not the entries along diagonal but only modulo signs and permutation. They are expressed in a manifestly gauge invariant fashion as
\be
\eta_a\equiv{\rm Tr}\,M^a, \, a=1,\ldots, \#\,{\rm cross\mbox{-}ratios}\qquad {\rm where}\quad M_{\alpha\beta}=(P_\alpha\cdot Q_\rho)(Q_\rho\cdot P_\beta).
\ee
Formally, the above definition of $\eta_a$ can be extended for $a>$($\#$ cross-ratio) but they all can be expressed in terms of the above, thanks to the Caley-Hamilton theorem or, more appropriately, trace relations.
We obtain a convenient graphical representation of cross-ratios, by representing the basic ``building blocks" as
\be
P_\alpha^A P_\alpha^B =:\,\,^A(PP)^B \sim \, A-\hspace{-0.15cm}-\hspace{-0.07cm}\CIRCLE\hspace{-0.13cm}-\hspace{-0.15cm}-B\qquad
Q_\rho^A Q_\rho^B =:\,\,^A(QQ)^B \sim \, A-\hspace{-0.15cm}-\hspace{-0.07cm}\Circle\hspace{-0.13cm}-\hspace{-0.15cm}-B.
\ee
Two dots (of any color) are connected when one of their vector indices is contracted. The orthonormal condition implies
\begin{center}
\includegraphics[scale=0.3]{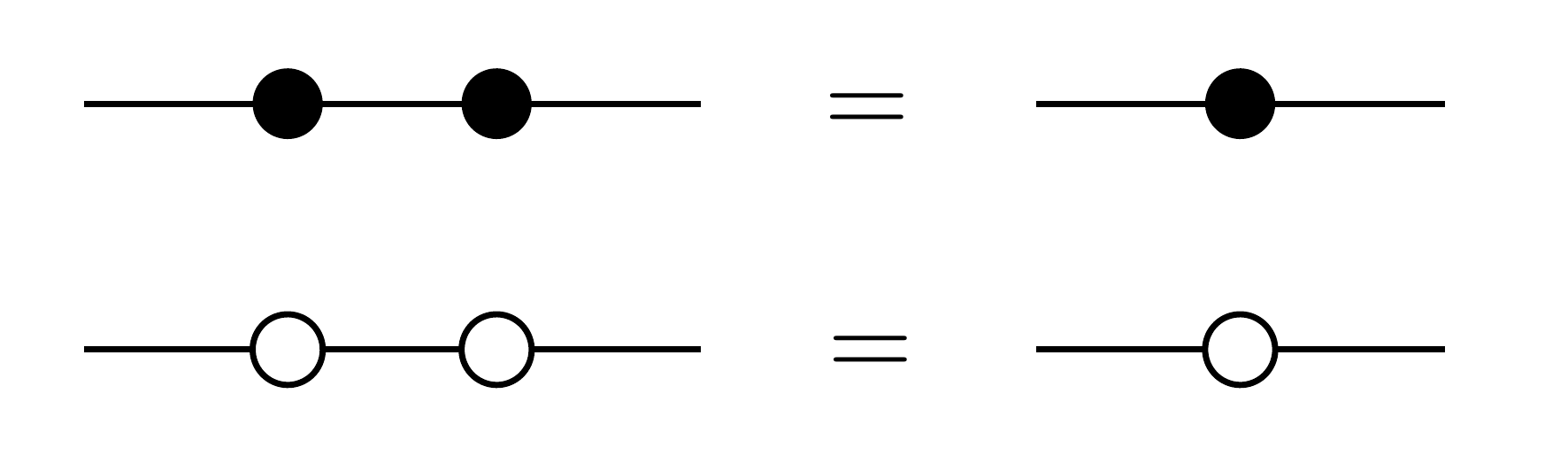}\qquad
\includegraphics[scale=0.3]{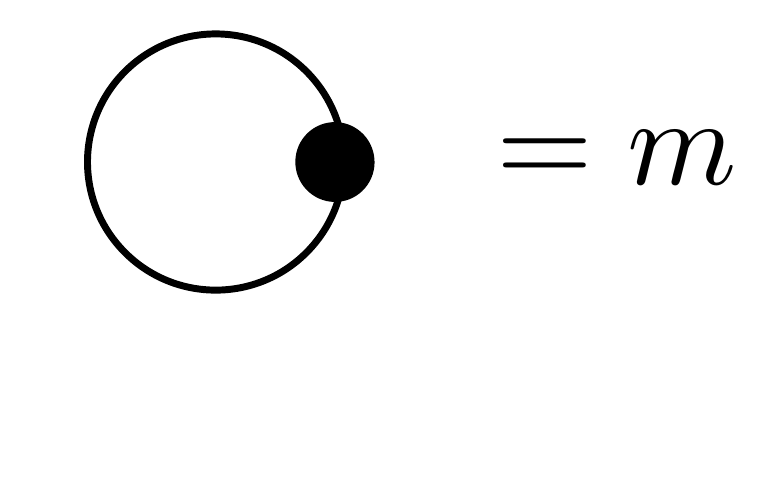}\qquad
\includegraphics[scale=0.3]{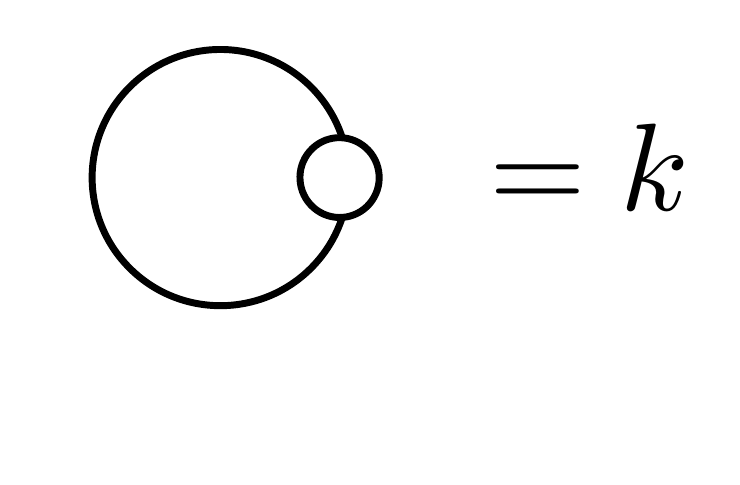}
\end{center}
The cross-ratio $\eta$ is represented as a closed chain with $a$ pairs of black and white dots.
\begin{center}
\includegraphics[scale=0.3]{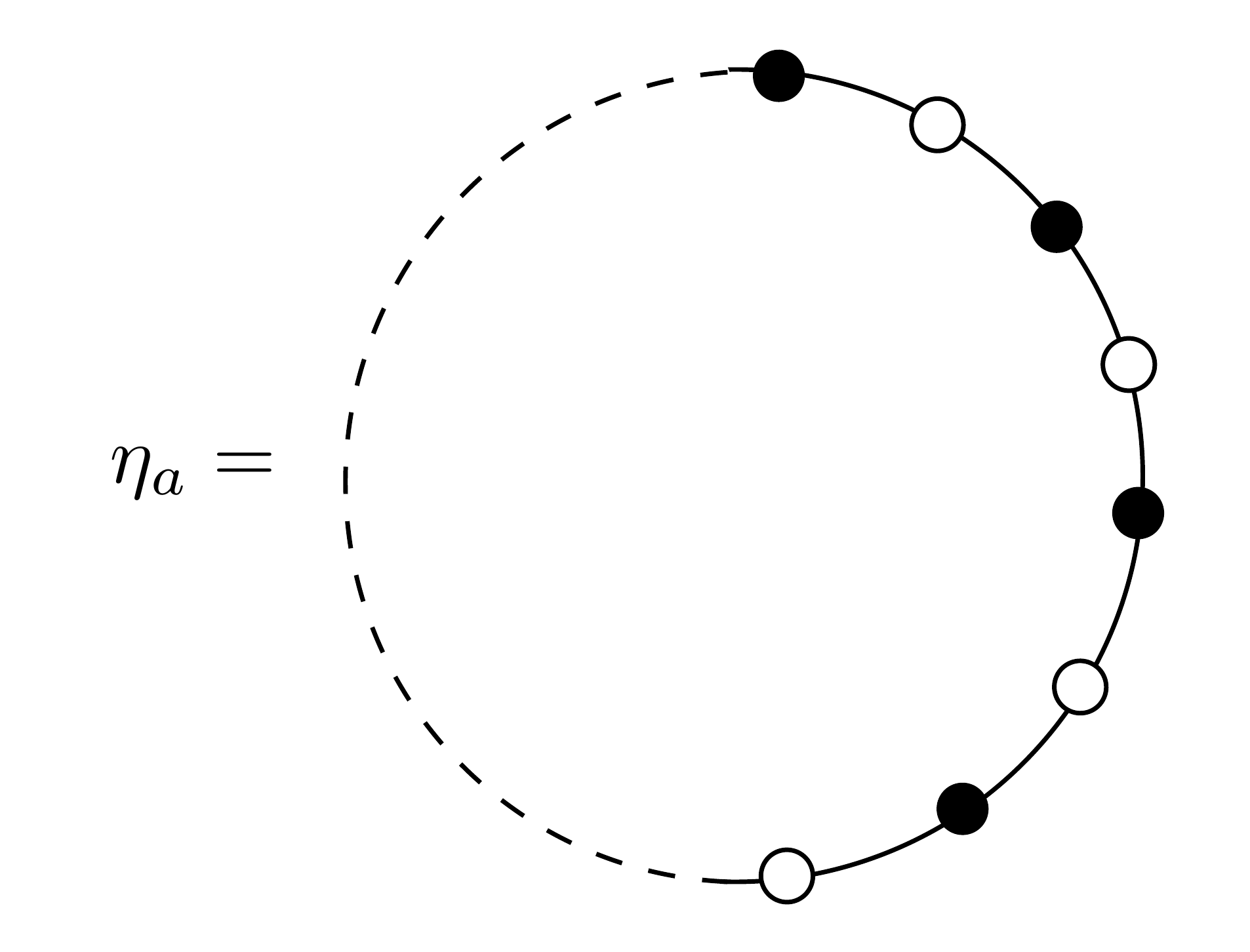}
\end{center}

After identifying the cross-ratios we now in a position to compute the two point function of defects. Conformal invariance restricts its form to
\be
\langle \CD^{(m)}(P_\alpha) \CD^{(k)}(Q_\rho)\rangle=
F(\eta_a)=\sum_{\Phi_{\Delta,R}} C^{\CD^{(m)}}_{\Phi} C^{\CD^{(k)}}_\Phi G_{\Delta,R}(\eta_a)
\ee
The undetermined function of the cross-ratios $F(\eta_a)$ gets a contribution from the  local operator $\Phi_{\Delta,R}$ appearing in the expansion of the defect $\CD^{(m)}(P_\alpha)$ (and $\CD^{(k)}(Q_\rho)$). Here $\Delta$ is the conformal dimension and $R$ is the representation under the rotational group $SO(d)$.
The contribution is proportional to the two defect expansion coefficients $C^{\CD^{(m)}}_\Phi$ and $C^{\CD^{(k)}}_\Phi$ and is multiplied by a kinematical function of cross-ratios known as the conformal block $G_{\Delta,R}(\eta_a)$. The conformal block  satisfies the eigenvalue equation,
\be
(L^2+C_{\Delta,R})\,G_{\Delta,\ell}(\eta_a)=0,\qquad
C_{\Delta,R}=\Delta(\Delta-d)+{\rm Cas}_{R}.
\ee
Here ${\rm Cas}_R$ is the eigenvalue of the quadratic Casimir of $SO(d)$ for representation $R$. For $l$ index traceless symmetric tensor, $C_R=l(l+d-2)$.
The operator  $L^2$ is the quadratic Casimir operator for the conformal group $SO(d+1,1)$.
\be
L^2=\frac12 L^{AB}L_{AB},\qquad L_{AB}=\sum_\alpha \Big(P_\alpha^{A}\frac{\partial}{P_\alpha^{B}}-P_\alpha^{B}\frac{\partial}{P_\alpha^{A}}\Big).
\ee
In what follows, we will obtain the differential equation obeyed by the conformal block in terms of conformal cross-ratios $\eta_a$ and solve them is some special cases, effectively reducing the computation of the defect two point function to that of the defect expansion coefficients.

In order to write the eigenvalue equation in terms of cross-ratios, we need $L^2\eta_a$ and $\frac12 L^{AB}\eta_a \, L_{AB} \eta_b$.
It is easy to figure out the action of $L_{AB}$ on the cross-ratio $\eta_a$ using its graphical representation. It essentially breaks one of the links in the chain.

\begin{center}
\includegraphics[scale=0.3]{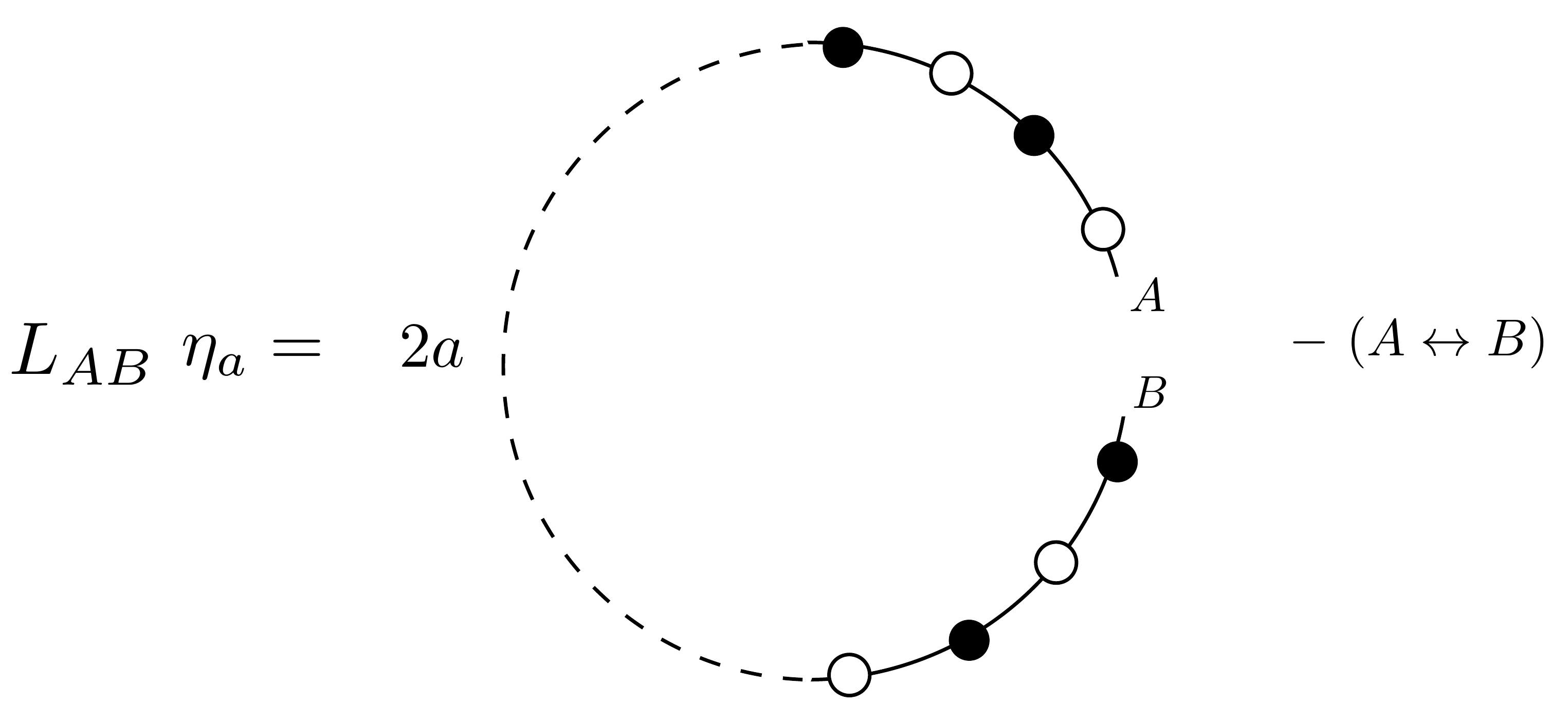}
\end{center}
From this we get,
\bea
\frac12 L^{AB}\eta_a \, L_{AB} \eta_b &=& 4ab (\eta_{a+b-1}-\eta_{a+b})\\
L^2 \eta_a&=&2a\Big[(a-1+m+k)\eta_{a-1}-(a+1+d)\eta_a+\sum_{b=1}^{a-2}\eta_{b} \eta_{a-1-b}-\sum_{b=1}^{a-1}\eta_{b} \eta_{a-b}\Big]\nonumber
\eea
It is understood that $\eta_a$'s with $a>$($\#$ cross-ratios) are expressed in terms of physical cross-ratios using the trace relations. Here, the second equation is valid for $a\geq 3$. For $a=1,2$,
\bea
L^2 \eta_1&=&2(mk-(d+2) \eta_1)\label{eta1casimir}
\\
L^2 \eta_2&=&4\Big[(1+m+k)\eta_1-(d+3)\eta_2-\eta_1^2\Big]\label{eta2casimir}
\eea
The Casimir eigenvalue equation becomes,
\be \label{casimir}
\sum_{a,b} \Big(\frac12 L^{AB}\eta_a \, L_{AB} \eta_b\Big) \frac{\partial^2}{\partial \eta_a\partial \eta_b} +\sum_a \Big(L^2 \eta_a\Big) \frac{\partial}{\partial \eta_a} +C_{\Delta,l}=0.
\ee

In the case $d<m+k-2$, it is more convenient to use ``dual" cross-ratios $\deta_i=Tr\dM^i$ where $\dM_{\alpha\beta}=(\dP_\alpha\cdot \dQ_\rho)(\dQ_\rho\cdot \dP_\beta)$. The above discussion remains valid for dual cross-ratios with the substitution $m\to d+2-m, k\to d+2-k$.

\subsection{Special cases}
In this subsection, we study the conformal blocks for two defect configuration in certain special cases. Given that the number of cross-ratios for co-dimension $m$ and co-dimension $k$ defect is min$(m,k,d+2-m,d+2-k)$, the simplest case, involving a single cross-ratio, is the correlation of co-dimension $1$ defect with a defect of arbitrary co-dimension $m$.

\subsubsection{Correlation with co-dimension $1$ defect}\label{codim1}
\begin{figure}[h]
\centering
\includegraphics[scale=0.4]{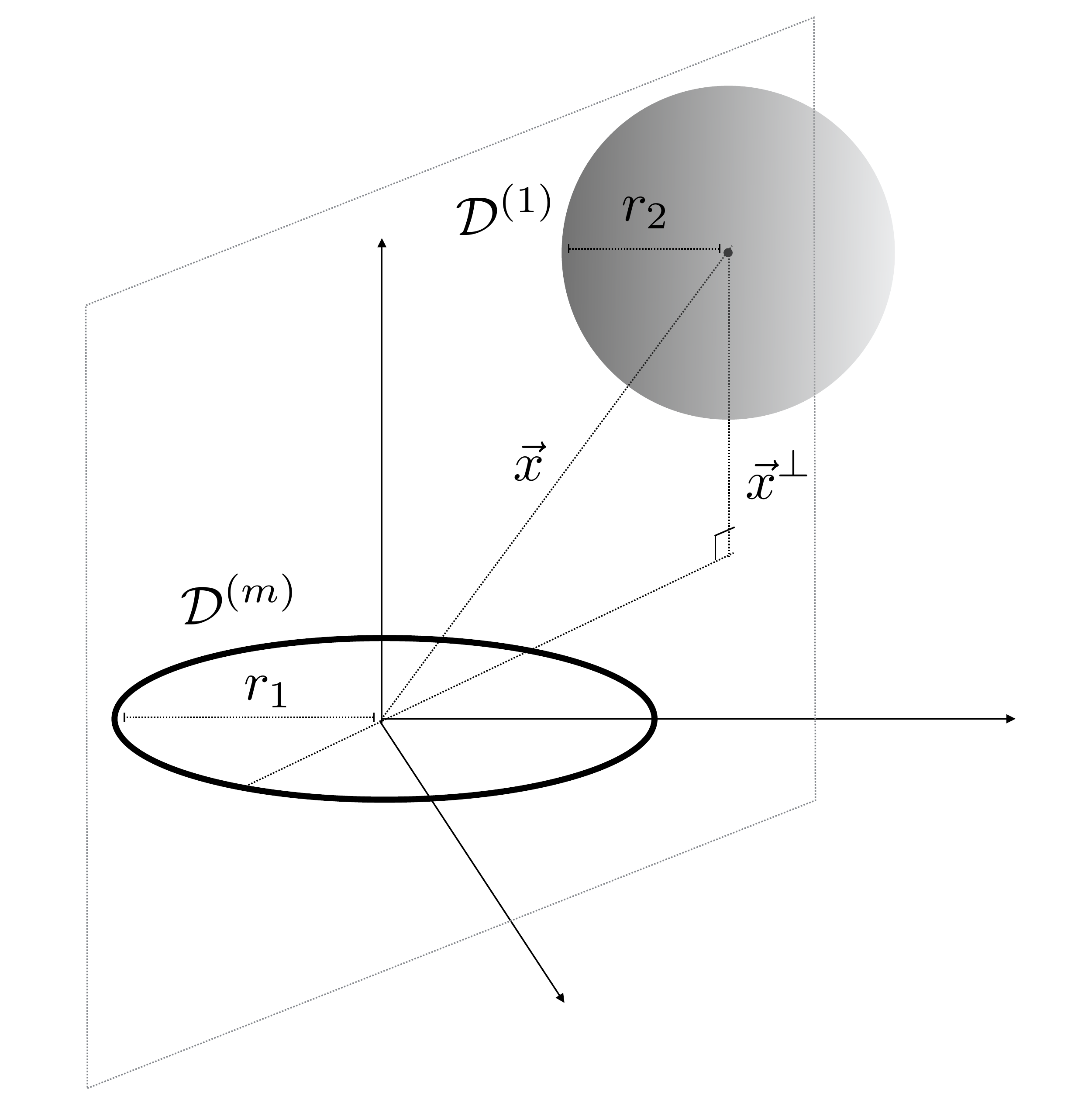}
\caption{A generic configuration of co-dimension $m$ and co-dimension $1$ defect.}
\label{codim1m}
\end{figure}

A generic configuration of  $\CD^{(m)}(P_\alpha)$ near  $\CD^{(1)}(Q)$ is depicted in figure \ref{codim1m}.  The former is of radius $r_1$, centered at the origin and is contained in the subspace spanned by $e_j,j=1,\ldots, d-m+1$. Its coordinates have been determined in equation \eqref{sphere-p}. The latter is of radius $r_2$ and is centered at $\vec x$. Its coordinates  are $Q=(\frac{1}{r_2},\frac{|\vec x|^2}{r_2}-r_2,\frac{|\vec x|}{r_2})$. The only cross-ratio $\eta_1$ is
\be
\eta_1=(P_{\alpha} \cdot Q)(Q\cdot P_{\alpha})=\frac{|\vec x^\perp|^2}{r_2^2}+(\frac{|\vec x|^2-r_1^2-r_2^2}{2 r_1 r_2})^2.
\ee
Here, $|\vec x^\perp|$ is the orthogonal distance of the center of $\CD^{(1)}$ from the subspace containing $\CD^{(m)}$. In the defect expansion limit, $r_1,r_2\to 0$ and the cross-ratio $\eta_1\to \infty$ while in the limit when the defects touch each other, $\eta_1\to 1$.
An interesting geometry is when $\CD^{(1)}$ becomes flat. This is achieved by taking $|\vec x|,r_2\to \infty$ with $s\equiv |\vec x|-r_2$ kept fixed. In this limit the cross-ratio simplifies,
\be\label{cross1m}
\eta_1=(\frac{|\vec x^\perp|}{|\vec x|})^2+(\frac{s}{r_1})^2.
\ee

A co-dimension one defect can have nonzero correlation only with a scalar local operator. Hence the eigenvalue of the conformal Casimir is $C_{\Delta,\cdot}=\Delta(\Delta-d)$.
In order to get the conformal block, we substitute $k=1$ in equation \eqref{eta1casimir}. The Casimir eigenvalues equation is
\be\label{casimireq1m}
(\eta_1-\eta_2)\frac{\partial^2}{\partial \eta_1\partial \eta_1}+\frac12(m-(d+2) \eta_1)\frac{\partial}{\partial \eta_1}+\frac14 \Delta(\Delta-d)=0.
\ee
The $\eta_2$ appearing in the above equation is related to $\eta_1$ as $\eta_2=\eta_1^2$. This is because $\eta_i={\rm Tr} M^i$ where $M$ is a $1\times 1$ matrix. We recognize the equation as the standard hypergeometric differential equation. In the limit $\eta_1\to \infty$, the leading behavior of the conformal block is determined by the primary term of the defect expansion. So, from \eqref{planar-bulk}, we expect the conformal block to go as $\eta_1^{-\frac{\Delta}{2}}$. This property, along with the eigenvalue equation \eqref{casimireq1m} fixes the conformal block.
\be\label{block1m}
G_{\Delta, \cdot}=\eta_1^{-\frac{\Delta}{2}} \,_2F_1(\frac{\Delta}{2},1+\frac{\Delta-m}{2},1+\Delta-\frac{d}{2};\eta_1^{-1}).
\ee

\subsection*{Two local operators near a co-dimension $1$ defect}

An interesting special case is when $m=d$. In this case,  $\CD^{(m)}$ is simply a pair of local operators with formal conformal dimension $0$. The relevant conformal block is 
\be\label{block1d}
G_{\Delta,\cdot}=\eta_1^{-\frac{\Delta}{2}} \,_2F_1(\frac{\Delta}{2},1+\frac{\Delta-d}{2},1+\Delta-\frac{d}{2};\eta_1^{-1}).
\ee

The conformal blocks in this case have been previously studied in the literature \cite{McAvity:1995zd, Liendo:2012hy}. There the role of the co-dimension $1$ defect is played by a conformal boundary. For a planar boundary and the local operator insertions at $\vec x_1$ and $\vec x_2$, the authors define a cross-ratio
\be
\xi=\frac{|\vec x_1-\vec x_2|^2}{4 x_1^\perp x_2^\perp}
\ee
where, $x^\perp$ is the perpendicular distance of the local operators from the boundary.  Their cross-ratio is related to ours as $\eta_1=1+1/\xi$. With this substitution and setting external operator conformal dimensions $\Delta_{1,2}=0$, their conformal block
\be
\hat G_{\Delta,\cdot}=\xi^{\frac{\Delta-(\Delta_1+\Delta_2)}{2}} \,_2F_1(\frac{\Delta+\Delta_1-\Delta_2}{2},\frac{\Delta-\Delta_1+\Delta_2}{2},\Delta-\frac{d}{2}+1;-\xi).
\ee
agrees with ours \eqref{block1d}.

\subsubsection{Two co-dimension $2$ defect}
\begin{figure}[h]
\centering
\includegraphics[scale=0.4]{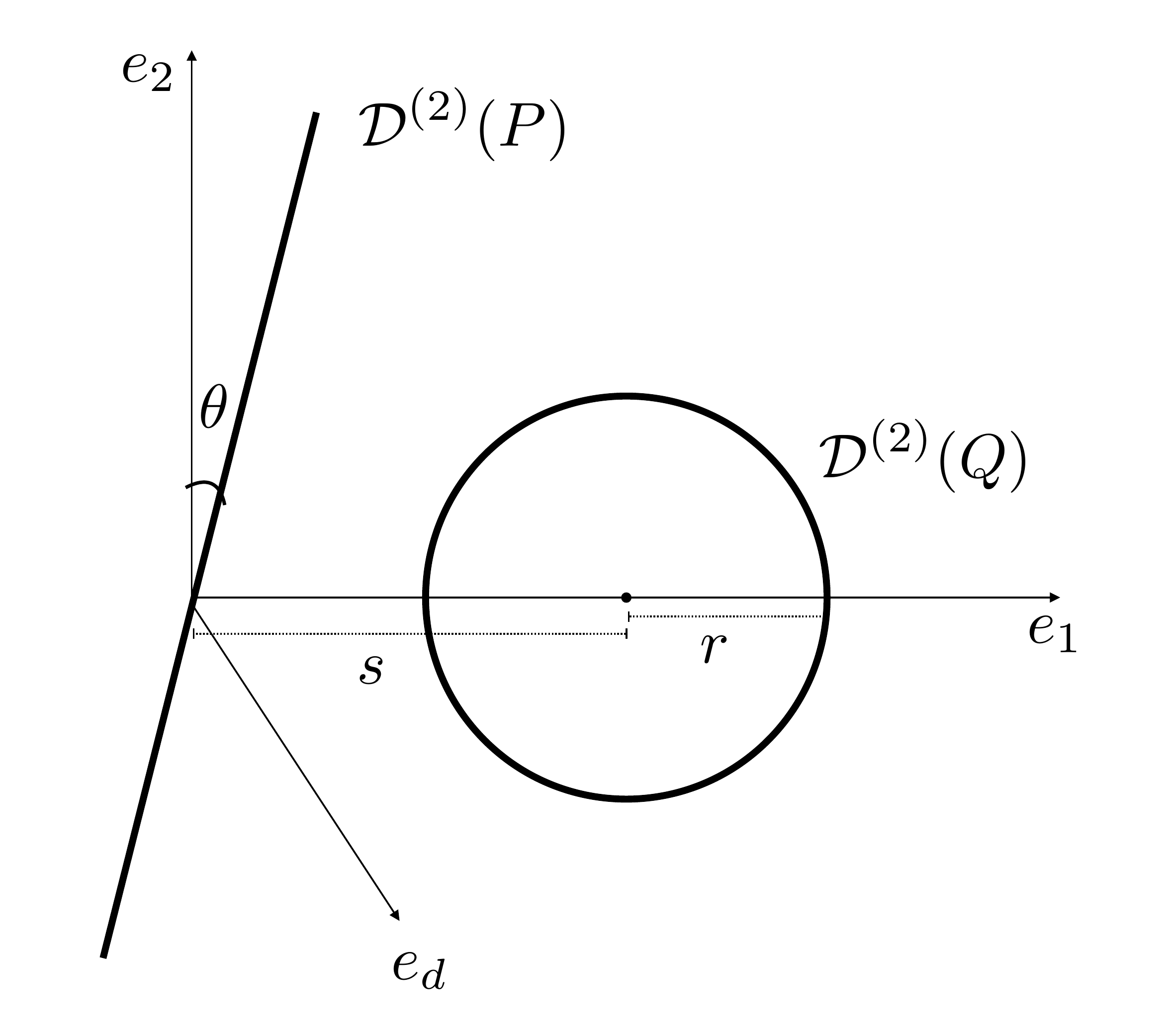}
\caption{A canonical configuration of two co-dimension $2$ defects. One of them is a sphere of radius $r$ sitting at $se_1$ in the hyperplane transverse to $e_d$. The other one is flat and is tilted by an angle $\theta$ in the $e_2-e_d$ plane.}
\label{codim22}
\end{figure}

Let us move on to the next simplest case, the of correlation function of two co-dimension $2$ defects $\CD^{(2)}(P_\alpha)$ with $\CD^{(2)}(Q_\mu)$. The two conformal cross-ratios are conveniently understood as follows. Take $\CD^{(2)}(P)$ to be flat and living in a plane spanned by $e_2,\ldots, e_{d-1}$. The other defect $\CD^{(2)}(Q)$ to be circular with radius $r$, centered at $s e_1$ and living in a plane spanned by $e_1,\ldots, e_d$. To obtain a general configuration, we tilt the flat defect in a $e_{2}-e_{d}$ plane by angle $\theta$. This geometry is shown in figure \ref{codim22}. The coordinates of these defects are calculated in equation \eqref{shifted-defect} and \eqref{tilted-defect} respectively. 
\bea
P_1=(0,0,e_1),&&\,P_2=(0,0,e_d\cos\theta-e_2\sin\theta)\nonumber\\
Q_1=(0,0,e_d),&&\,Q_2=(\frac1r,-r+\frac{s^2}{r},\frac{s}{r}e_1).
\eea
The matrix of dot products is
\be
P_\alpha\cdot Q_\mu=\Big(
\begin{array}{cc}
0&\cos\theta\\
\frac{s}{r}&0
\end{array}
\Big),\quad\qquad\,\,M\equiv{(P_\alpha\cdot Q_\mu)( Q_\mu \cdot P_\beta})=
\Big(
\begin{array}{cc}
\cos^2\theta&0\\
0&(\frac{s}{r})^2
\end{array}
\Big).
\ee
The cross-ratios $\eta_1$ and $\eta_2$ are
\be
\eta_1=\cos^2\theta+(\frac{s}{r})^2,\qquad \eta_2=\cos^4\theta+(\frac{s}{r})^4.
\ee
From here it is clear that the defect expansion limit is $\eta_{1,2}\to \infty$. 
The trace relations for a $2\times 2$ matrix are $\eta_3=\frac32 \eta_1\eta_2-\frac12 \eta_1^3, \eta_4=\frac12 (\eta_2^2-\eta_1^4)+\eta_2\eta_1^2$. With these substitutions, the conformal Casimir equation becomes
\bea
&&(\eta_1-\eta_2)\frac{\partial^2}{\partial \eta_1^2}+2(\eta_1^3+2\eta_2-3\eta_1\eta_2)\frac{\partial^2}{\partial \eta_1\partial\eta_2}+2(\eta_1^2-\eta_2)(\eta_1^2-\eta_1+\eta_2)\frac{\partial^2}{\partial\eta_2^2}\nonumber\\
&&(2-\frac{d+2}{2}\eta_1)\frac{\partial}{\partial\eta_1}+(5\eta_1-(d+3)\eta_2-\eta_1^2)\frac{\partial}{\partial \eta_2}+C_{\Delta,R}=0. \label{codim2eq}
\eea
Because only the symmetric traceless tensors can appear in the OPE, the eigenvalue of the Casimir is $C_{\Delta,R}=C_{\Delta,l}=\Delta(\Delta-d)+l(l+d-2)$.
This seemingly unfamiliar equation can be brought to a familiar form by a change of variables. We obtain the right variables by ``dualization". Consider the case of two dimension $0$ defects instead \emph{i.e.} of four local operators and study their conformal block. 

\subsection*{Four local operators}
In out set up, four local operators (of conformal dimension $0$) are modeled as two co-dimension $d$ defects. Instead of using $d$-dimensional hyperplanes in the embedding space to parametrize the defect, we use the orthogonal $2$-dimensional hyperplanes. These planes are spanned by the orthonormal frames $\dP$ and $\dQ$. The dual cross-ratios $\deta_1$ and $\deta_2$ obey the same exact equation as \eqref{codim2eq}.  

On the other hand, for four local operators, it is more common to use the cross-ratios $u$ and $v$ defined as 
\be\label{uv}
u=\frac{X_1\cdot X_2 \,X_3\cdot X_4}{X_1\cdot X_3 \,X_2\cdot X_4},\qquad v=\frac{X_1\cdot X_4 \,X_2\cdot X_3}{X_1\cdot X_3 \,X_2\cdot X_4}, 
\ee
where $X_i$ are the null-vectors in the embedding space corresponding to the position of $i$th local operator. Dolan and Osborn solved the conformal Casimir equation in this case and obtained the conformal block in terms of a product of two hypergeometric functions \cite{Dolan:2003hv}.
We take the dual frame $\dP$ to parametrize the defect consisting of points at $X_1$ and $X_2$. The other defect consists of points $X_3$ and $X_4$ and is parametrized by the dual frame $\dQ$. It is easy to relate these coordinates,
\bea
X_1&=&\dP_{\dplus}\equiv \dP_1+\dP_2, \qquad  \,\, X_2=\dP_{\dminus}\equiv \dP_1-\dP_2 \nonumber \\
X_3&=&\dQ_{\dplus}\equiv \dQ_1+\dQ_2, \qquad  X_4=\dQ_{\dminus}\equiv \dQ_1-\dQ_2.
\eea
Here $\dP_{\dpm}$ and $\dQ_{\dpm}$ are light-cone directions in the planes parametrized by $\dP$ and $\dQ$ respectively.
Substituting in \eqref{uv},
\be
u=\frac{4}{\dP_\dplus \cdot \dQ_\dplus \, \dP_\dminus \cdot \dQ_\dminus},\qquad 
v=\frac{\dP_\dplus \cdot \dQ_\dminus \, \dP_\dminus \cdot \dQ_\dplus}{\dP_\dplus \cdot \dQ_\dplus \, \dP_\dminus \cdot \dQ_\dminus}.
\ee
Now we are ready to express the new cross-ratios $\deta$ in terms the old $u,v$.
\bea
\deta_1 &=& (\dP_\dalpha \cdot \dQ_\dmu) (\dQ^\dmu \cdot \dP^\alpha)=\frac{2(1+v)}{u}\nonumber\\
\deta_2 &=& (\dP_\dalpha \cdot \dQ_\dmu) (\dQ^\dmu \cdot \dP^\dbeta) (\dP_\dbeta \cdot \dQ_\dnu) (\dQ^\dnu \cdot \dP^\dalpha)=\frac{2(1+6v+v^2)}{u^2}.
\eea
Remarkably, after substituting this change of variables in \eqref{codim2eq} for $\deta$'s, we recover the differential equation for the  usual four point function conformal block in terms of the more conventional cross-ratios $u$ and $v$. 
Because, $\eta_{1,2}$ obey the same equation as $\deta_{1,2}$, the conformal blocks of \cite{Dolan:2003hv} straightforwardly carry over for co-dimension $2$ defects as well. For completeness, we reproduce their result here.
\bea
G_{\Delta,l}&=&(-1)^l \frac{x z}{x-z}(k_{\Delta+l}(x)k_{\Delta-l-2}(z)-(x\leftrightarrow z))\nonumber\\
k_\beta(x) &=& x^{\beta/2} {_2F_1}(\frac{\beta}{2},\frac{\beta}{2},\beta;x)
\eea
where the variables $x,z$ are related to $\eta_{1,2}$ as,
\be
\eta_1=\frac{2(1+v)}{u}|_{u=xz,v=(1-x)(1-z)},\qquad \eta_2=\frac{2(1+6v+v^2)}{u^2}|_{u=xz,v=(1-x)(1-z)}.
\ee
From \eqref{planar-bulk}, for scalar operators, we expect the conformal block to go as $\eta_1^{-\frac{\Delta}{2}}\eta_1^{-\frac{\Delta}{4}}$ in the defect expansion limit $\eta_{1,2}\to \infty$. This is consistent with the above solution.
It is interesting to study the conformal Casimir equation for general co-dimension defects.

\section{Discussion}\label{examples}
In this paper we have studied the constraints imposed by conformal invariance on the correlators of nonlocal operators. In the case of local operators, it has long been on known that their correlation functions can be fixed from the knowledge of the operator product expansion coefficients. In the case of defects also, we have shown that problem of computing correlation functions reduces to the computation of defect expansion coefficients. To realize the usefulness of this formalism, it would be desirable to compute the defect expansion coefficients explicitly in some examples. Even a free conformal  scalar field theory makes for an interesting example. We expect the computation to have a straightforward generalization to the Maxwell theory as well as to the Wilson-Fisher fixed point. If we interpret the free energy of the two defect system as potential, the correlation gives rise to a force. 
In the case of Maxwell theory, this is the celebrated Casimir force. For general CFTs, the resulting force is known as the critical Casimir force. It has been studied in experimentally \cite{naturecritical} as well as numerically \cite{0295-5075-80-6-60009} for CFTs in the universality class of $3d$ Ising model for spherical co-dimension $1$ defects. 
Our work should provide its exact dependence on the relative geometry of the defects in question. 
Unlike the usual Casimir force, by a suitable engineering of defects, the critical Casimir force can be tuned from being attractive to being repulsive and vice versa, due to this flexibility, the critical Casimir force is expected to have applications in the construction of \emph{micro electro-mechanical systems (MEMS)} \cite{refId0}. 

Defect conformal blocks can also be applied to compute the Renyi entropy for two or more spatial regions in a CFT\footnote{We thank Stefan Leichenauer for suggesting this possibility.}. This involves computing correlation of co-dimension $2$ twist operators which incorporate the replica trick. We have seen that, in even dimensions, the conformal Casimir equation can be solved in closed form to obtain the conformal blocks as a product of two hypergeometric functions. It would then be very instructive to work out twist defect expansion coefficients in free or holographic CFTs.

Our work has natural generalization in multiple directions. 
In our discussion, we have focused on correlations of defects when they do not carry any spin and do not carry insertions of defect local operators. Generalization to the later case should be useful in 
computing correlations of arbitrarily shaped defects by taking the defect local operator to be the displacement operator $D^\alpha$ of \eqref{displacement}.  Another generalization is to the Lorentzian CFTs. Because, a ``sphere" in the Lorentzian theory can be of three types depending on whether the radius $r>0,r<0$ or $r=0$, the conformal defects are classified accordingly. It would be interesting to investigate constraints of causality along with those imposed by conformal invariance.
Another generalization is to the supersymmetric theories. For special configurations of defects it should be possible to compute the correlation via localization. For general co-dimension defects, equation \eqref{casimir} for the quadratic Casimir and similar eigenvalue equations for higher Casimir are mathematically interesting. 
Recently, in \cite{Isachenkov:2016gim} the authors observe a connection of the conformal Casimir equation with the integrable Hamiltonian of the Calogero-Sutherland model. We expect the defect conformal blocks only to enrich this connection.

When the co-dimensions $m$ and $k$ of the two defects satisfy $m+k=d-1$, the defects can link. In this configuration, the defect expansion as presented in section \ref{defectexp} is not valid. We have to consider another expansion where the quantization surface enclosing one defect cuts the other. The states are expanded in terms of the defect local operators on the latter defect. It would be nice to work out this case in detail. Finally, the ultimate goal would be to  obtain a crossing equation for two point function of defects analogous to the crossing equation for four point function of local operators. In the case of local operators, the crossing equation has been used with a great success in the \emph{conformal bootstrap} program to constrain the space of CFTs. Having a version for defects would allow one to put constraints on the spectrum of defects. Of course, there are obvious hurdles in obtaining such an equation. We suspect one such problem is establishing a state/operator correspondence for defects.
Clearly, the computation of defect conformal blocks opens doors to many new unexplored directions. We wish to pursue them in the future.

\section*{Acknowledgements}
We thank Chris Beem, Yu Nakayama, Leonardo Rastelli and Nathan Seiberg for useful comments. We would especially like to thank  David Simmons-Duffin for stimulating discussions and for collaboration in the early part of this work. We also thank the hospitality of International Center for Theoretical Sciences, Bangalore where a part of this work was carried out. The author's research is supported by the Roger Dashen Membership Fund and the National Science Foundation grant PHY-1314311.

\bibliographystyle{JHEP}
\bibliography{defects}
\end{document}